\documentclass[12pt,epsf]{article}
\usepackage{graphicx}
\usepackage{amsmath}
\usepackage{amsfonts}
\begin{document}

\def\a{\alpha}
\def\b{\beta}
\def\c{\varepsilon}
\def\d{\delta}
\def\e{\epsilon}
\def\f{\phi}
\def\g{\gamma}
\def\h{\theta}
\def\k{\kappa}
\def\l{\lambda}
\def\m{\mu}
\def\n{\nu}
\def\p{\psi}
\def\q{\partial}
\def\r{\rho}
\def\s{\sigma}
\def\t{\tau}
\def\u{\upsilon}
\def\v{\varphi}
\def\w{\omega}
\def\x{\xi}
\def\y{\eta}
\def\z{\zeta}
\def\D{\Delta}
\def\G{\Gamma}
\def\H{\Theta}
\def\L{\Lambda}
\def\F{\Phi}
\def\P{\Psi}
\def\S{\Sigma}

\def\o{\over}
\def\beq{\begin{eqnarray}}
\def\eeq{\end{eqnarray}}
\newcommand{\gsim}{ \mathop{}_{\textstyle \sim}^{\textstyle >} }
\newcommand{\lsim}{ \mathop{}_{\textstyle \sim}^{\textstyle <} }
\newcommand{\vev}[1]{ \left\langle {#1} \right\rangle }
\newcommand{\bra}[1]{ \langle {#1} | }
\newcommand{\ket}[1]{ | {#1} \rangle }
\newcommand{\EV}{ {\rm eV} }
\newcommand{\KEV}{ {\rm keV} }
\newcommand{\MEV}{ {\rm MeV} }
\newcommand{\GEV}{ {\rm GeV} }
\newcommand{\TEV}{ {\rm TeV} }
\def\diag{\mathop{\rm diag}\nolimits}
\def\Spin{\mathop{\rm Spin}}
\def\SO{\mathop{\rm SO}}
\def\O{\mathop{\rm O}}
\def\SU{\mathop{\rm SU}}
\def\U{\mathop{\rm U}}
\def\Sp{\mathop{\rm Sp}}
\def\SL{\mathop{\rm SL}}
\def\tr{\mathop{\rm tr}}

\def\IJMP{Int.~J.~Mod.~Phys. }
\def\MPL{Mod.~Phys.~Lett. }
\def\NP{Nucl.~Phys. }
\def\PL{Phys.~Lett. }
\def\PR{Phys.~Rev. }
\def\PRL{Phys.~Rev.~Lett. }
\def\PTP{Prog.~Theor.~Phys. }
\def\ZP{Z.~Phys. }


\baselineskip 0.65cm

\begin{titlepage}

\begin{flushright}
UT-10-10\\
IPMU-10-0099
\end{flushright}

\vskip 1.35cm
\begin{center}
{\large \bf
A Conformal Gauge Mediation and Dark Matter \\ with Only One Parameter}
\vskip 1.2cm
Tsutomu T. Yanagida and Kazuya Yonekura 
\vskip 0.4cm

{\it $^1$ Institute for the Physics and Mathematics of 
the Universe (IPMU),\\  
University of Tokyo, Chiba 277-8568, Japan\\ 
$^2$  Department of Physics, University of Tokyo,\\
    Tokyo 113-0033, Japan}

\vskip 1.5cm

\abstract{
If the supersymmetry (SUSY) is a solution to the hierarchy problem, it is puzzling that any SUSY particle has not been discovered yet. 
We show that there is a low-scale conformal gauge mediation model which contains all necessary 
ingredients, i.e. not only a SUSY-breaking dynamics and a gauge mediation mechanism, but also a candidate for the dark matter. 
The model has only one
free parameter, that is, the mass for messengers. 
In this model, the dark matter is provided by a composite particle in the SUSY-breaking sector, 
and the observed value of the dark matter
density uniquely fixes the mass of messengers at the order of $10^2$~TeV. 
Then, the sfermion and gaugino masses are fixed to be of order $10^2 \sim 10^3~\GEV$ without any arbitrariness, thus  
the SUSY particles are expected not to be discovered at the Tevatron or LEP, while having a discovery possibility at the LHC.
}
\end{center}
\end{titlepage}

\setcounter{page}{2}

\section{Introduction}

If the supersymmetry (SUSY) is a solution to the so-called hierarchy problem between the electroweak and the grand unified theory (GUT) scales, it is very much puzzling that the SUSY has not been discovered, yet. Thus, we need an answer to the question why  it has not been discovered. We discuss 
that a conformal gauge mediation~\cite{Yanagida:2010wf} based on the recently discovered SUSY breaking mechanism~\cite{Izawa:2009nz} has only one 
free parameter. This free parameter is, however, determined if a composite particle in the SUSY-breaking hidden sector is the dominant component of the dark matter (DM) in the universe. Therefore, the model does not have any free parameter
in determining the sfermion and gaugino masses of the minimal SUSY standard model (MSSM), and hence it may provide an answer to the question.

Let us first consider the gravity mediation. The lightest SUSY particle (LSP) can be the DM in the universe. We see that there are large parameter regions (with light squarks and sleptons of masses 10$\sim$100 GeV) consistent with both the electroweak symmetry breaking and the observed density of DM~\cite{Ellis:2010kf}. Thus, it seems hopeless to explain undiscovery of the SUSY particles at Tevatron or LEP experiments. In the anomaly mediation,
the wino is most likely the DM.  
If it is the case and if the winos were in the thermal bath in the early universe, the wino mass should be about 3 TeV to explain the DM density \cite{Hisano:2006nn}. Then, we can explain why the SUSY has not been discovered, yet, since all the masses of SUSY particles are above 3 TeV. However, the above argument has a loophole. That is, even the light wino of mass say 10 GeV can be the dominant DM if it is produced by gravitino decays and the reheating temperature is $T_R\simeq 10^{11}$ GeV \cite{Ibe:2004tg}. Thus, we can not, again, answer to the above question. 

Now we discuss the gauge mediation~\cite{Dine:1981za,mGMSB}. There are mainly two possible candidates for the DM in gauge mediation models. One is the gravitino, and the other  is a stable particle in the SUSY-breaking/messenger sector \cite{Dimopoulos:1996gy,Han:1997wn, Hamaguchi:2007rb,Ibe:2009dx,Shih:2009he,Fan:2010is}. In the case of the gravitino DM, the DM (gravitino) density in the present universe depends on the reheating temperature $T_R$ and hence we can not determined the SUSY-breaking scale. 

In the case of the SUSY-breaking/messenger sector DM, the mass of the DM may be determined if the hidden sector is strongly coupled~\cite{Dimopoulos:1996gy,Hamaguchi:2007rb,Fan:2010is} and the annihilation cross section of the DM is saturated by the unitarity bound \cite{Griest:1989wd}. Thus, one may determine the SUSY-breaking scale or the messenger scale depending on whether the DM is provided by the SUSY breaking sector or the messenger sector. 
However, the other mass scale of the messenger/SUSY-breaking sector which does not contain the DM is still undetermined and hence the masses of the MSSM particles are free parameters and can be in the range of $10\sim100$ GeV. Therefore, we have no reason why squarks and sleptons, for example, have not been discovered at Tevatron or LEP experiments. 

Interestingly, however, there is a class of gauge mediation models, called as conformal gauge mediation \cite{Ibe:2007wp}, where the SUSY-breaking scale is strongly linked to the messenger masses. In those models, all dimensionless parameters in the hidden sector are uniquely fixed at the infrared conformal 
fixed points, and the models are strongly coupled in phenomenologically interesting ways.
Those models have only one free parameter, that is the messenger mass, 
and if the hidden sector provides a DM candidate, the messenger mass can be fixed by the 
requirement that the DM density be consistent with the observed value.  
Thus, there is no arbitrariness in determining the sfermion and gaugino masses in the MSSM, 
and we find that they are too heavy to be observed at Tevatron or LEP experiments.
Therefore, those models may provide an answer to the above question~\footnote{We do not claim that 
the so-called little hierarchy problem is solved, but only try to explain the undiscovery of SUSY particles up to the energy scale $10\sim100$~GeV.}.
We should stress that the predicted masses for SUSY particles are in the region accessible to LHC experiments.

\section{Merits of strongly coupled one-parameter models} \label{sec:2}

Let us consider general strongly coupled models which have only one mass scale $\L$ as a free parameter.
Properties of such models may be analyzed by using a naive dimensional analysis~\cite{Luty:1997fk}.
In the naive dimensional analysis, the effective Lagrangian in the models with some fields $\f$ is estimated as
\beq
{\cal L} \sim \frac{1}{g_s^2} \L^4 \hat{\cal L}(\f/\L, \q_\mu \f/\L^2,\cdots),
\eeq
where $g_s \sim 4\pi$ represents strong coupling of the model, and $\hat{\cal L}$ is a dimensionless function with ${\cal O}(1)$ coefficients.
We neglect factors of color or flavor numbers $N$ in this section for simplicity.
After rescaling  $\f /g_s \to \f$ so that the kinetic terms are canonically normalized, one can see that the masses of particles are of order $\L$, 
Yukawa couplings are of order $g_s$, and the vacuum energy (in a global SUSY limit) is estimated as $V \sim \L^4/g_s^2$. 
 The DM mass if it exists is estimated as 
 \beq
 m_{\rm DM} \sim \L. \label{eq:DMmass}
 \eeq
 If the models are used for gauge mediation, then the sfermion and gaugino masses
 are of order
 \beq
 M_{\rm gaugino} \sim m_{\rm sfermion} \sim \frac{g_{\rm SM}^2}{16\pi^2}\L, \label{eq:sparticlemass}
 \eeq
where $g_{\rm SM}$ collectively denote the standard model (SM) gauge coupling constants. This may be determined so that these sparticle masses would become of order $\L$ if the SM gauge couplings $g_{\rm SM}$ were of order $4\pi$, and the power of $g_{\rm SM}$ is known 
from the general gauge mediation~\cite{Meade:2008wd}. The gravitino mass is of order
\beq
m_{3/2} \sim \frac{\sqrt{V}}{M_{Pl}} \sim \frac{\L^2}{g_s M_{Pl}},  \label{eq:gravitinomass}
\eeq
where $M_{Pl} \simeq 2.4\times 10^{18}~\GEV$ is the reduced Planck scale.

It is known~\cite{Griest:1989wd} that the thermal relic abundance of the strongly coupled DM becomes of order $\Omega h^2 \sim 0.1$ if the DM mass is of order
$m_{\rm DM} \sim {\cal O}(100)~\TEV$. From Eqs.~(\ref{eq:DMmass},\ref{eq:sparticlemass},\ref{eq:gravitinomass}), we obtain the sparticle masses of order
\beq
 M_{\rm gaugino} \sim m_{\rm sfermion} &\sim& {\cal O}(10^2 \sim 10^3)~\GEV,\\
 m_{3/2} &\sim& {\cal O}(1)~\EV.
\eeq
As discussed in the introduction, if the sfermions and gauginos in the MSSM have masses of order 
${\cal O}(10^2 \sim 10^3)~\GEV$, they are not neccesarily discovered at 
the Tevatron or LEP. Furthermore, the gravitino mass is very light. A gravitino mass less than $16~\EV$ is favored~\cite{Viel:2005qj},
because otherwise the model suffers from severe cosmological problems caused by the gravitino.
Thus, the strongly coupled one-parameter models give the desired order of masses for all relevant particles, i.e. 
the DM, the sfermions, the gauginos, and the gravitino.

There are several models of strongly coupled hidden sector with the above mass spectrum~\cite{Dimopoulos:1996gy,Hamaguchi:2007rb,Fan:2010is}.
These models are phenomenologically successful, but they have messenger mass scales and SUSY breaking scales 
as independent parameters, unfortunately.  
Then, one has to tune those mass scales by hand to achieve the above scenario, so we encounter a fine-tuning problem in the hidden sector~\footnote{
In many models of non-minimal gauge mediation, the fine-tuning is also required to avoid a splitting
between the sfermion masses and the gaugino masses. See e.g. Refs.~\cite{Dumitrescu:2010ha}, \cite{Shirai:2010rr} and references therein.}.

Conformal gauge mediation~\cite{Ibe:2007wp} addresses the above problem by linking the messenger mass to the SUSY-breaking scale.
In this scenario, the models are on strongly coupled infrared conformal fixed points above the mass of the messengers. 
After the decoupling of the messenger fields,
the hidden-sector gauge coupling becomes strong and then the SUSY is dynamically broken. If the infrared fixed point is very strongly coupled,
the hidden-sector gauge coupling blows up quickly after the decoupling of the messengers, so the messenger mass and the SUSY breaking scale
is almost the same.     
Thus, the scenario discussed in this section works naturally in the strongly coupled conformal gauge mediation.

If the mass of the messengers were generated by confining dynamics of gauge theory,
the above models would really be a theoretically and phenomenologically excellent realization of dynamical SUSY breaking and direct gauge mediation 
envisioned in Refs.~\cite{Witten:1981nf,Affleck:1984xz}.
Unfortunately, we have to put the messenger mass by hand in the conformal gauge mediation. However, it may be possible to retrofit the models~\cite{Dine:2006gm} so that the messenger mass is generated dynamically (by using gaugino condensation or something). 
Then the models may become fully satisfactory ones.
In this paper we do not attempt to find a model for messenger mass generation, but simply assume its existence.

\section{Model}

\subsection{Conformal gauge mediation}
In this subsection, we review the strongly coupled conformal gauge mediation model~\cite{Yanagida:2010wf} which is based on 
the  recently discovered dynamical SUSY-breaking mechanism~\cite{Izawa:2009nz}.
We consider a SUSY $SU(N_C)$ gauge theory with $N_Q$ vector-like pairs of quarks $Q^i$ and anti-quarks ${\tilde Q}_{\tilde j}$ ($i,{\tilde j}=1,\cdots,N_Q$). They belong to fundamental and anti-fundamental representations of the $SU(N_C)$, respectively. Here, we have omitted the $SU(N_C)$ color indices. We introduce $N_Q \times N_Q$ gauge singlet chiral multiplets $S^{\tilde i}_{~j}$. We choose $N_Q < N_C$, then we have a runaway type of 
dynamical superpotential at low energies~\cite{Affleck:1983mk}. 
In addition to the above fields we introduce $N_P$ flavors of massive quarks $P^a$ and ${\tilde P}_a$ which transform also as fundamental and anti-fundamental representations of the hidden gauge group $SU(N_C)$, respectively. Here, $a=1,\cdots,N_P$. The tree level superpotential of the model is given by
\begin{equation}
W=\lambda S^{\tilde j}_{~i}Q^i{\tilde Q}_{\tilde j} + m_P P^a{\tilde P}_a. \label{eq:treesuper}
\end{equation} 
The massive quarks play two roles. One is to stop the runaway of the potential and generate  SUSY breaking vacua \cite{Izawa:2009nz} and the other is to act as messengers of the SUSY breaking \cite{Yanagida:2010wf}. For the latter purpose we take $N_P=5$ and embed the GUT $SU(5)_{\rm GUT}$ 
into the flavor symmetry $SU(N_P)$ acting on $P$ and ${\tilde P}$. 
We see that our model belongs to  so-called semi-direct gauge mediation~\cite{Izawa:1997hu,Seiberg:2008qj}.
For the former purpose we restrict our discussion to a conformal window $\frac{3}{2}N_C < N_Q+N_P < 3N_C$ 
of the model~\cite{Seiberg:1994pq,Barnes:2004jj}. 
We assume that the model is nearly on the infrared fixed point at the scale above the threshold of the massive quarks $P,\tilde{P}$. The reason why we consider the conformal theory becomes clear below. 

The above superpotential has a flavor symmetry $SU(N_Q)_L \times SU(N_Q)_R$ which acts on the indices $i$ and ${\tilde j}$, respectively.
However, we do not impose this flavor symmetry on the model.
The form of the superpotential $W \sim SQ\tilde{Q}+P{\tilde P}$ may be ensured by some discrete R symmetry as we will see later,
and then the common Yukawa coupling $\l$ in the superpotential Eq.~(\ref{eq:treesuper}) may be achieved approximately by 
the infrared stability of the fixed point~\cite{Luty:2001jh}.

Let us discuss the dynamics of the SUSY breaking in the present model. The details can be found in Refs.~\cite{Izawa:2009nz,Yanagida:2010wf}
and we give only a rough sketch here.
At the classical level, the moduli space of the model is spanned by the vacuum expectation value (vev) of $S^{\tilde j}_{~i}$,
since the equations of motion make the vevs of quarks $Q,\tilde{Q},P$ and $\tilde{P}$ to be zero.
For simplicity, let us consider the direction $S^{\tilde j}_{~i}=S \d^{\tilde j}_{~i}$.
Then, the vev of $S$ gives masses to the quarks $Q,\tilde{Q}$. After integrating out all the quarks, 
we obtain a dynamical superpotential~\cite{Affleck:1983mk,Affleck:1984xz}
\beq
W_{\rm eff} \sim S^{\frac{N_Q}{N_C}}.\label{eq:effsuperp}
\eeq
This is a runaway superpotential when $N_Q<N_C$.
On the other hand, above the mass scale of the quarks $P,\tilde{P}$, the model becomes a superconformal theory~\cite{Seiberg:1994pq,Barnes:2004jj}.
Then, the superconformal symmetry (which is spontaneously broken by the vev of $S$) indicates that the low energy effective K\"ahler potential is given by
\beq
K_{\rm eff} \sim (S^\dagger S)^{\frac{1}{\D_S}}+\cdots, \label{eq:effkahlerp}
\eeq
where $\D_S$ is the scaling dimension of $S$ at the fixed point, and dots denote terms depending on $m_P$, the explicit breaking of superconformal symmetry,
which are negligible in the limit $m_P/S \to 0$. Note that the above K\"ahler potential has a scaling dimension 2 due to the power $1/\D_S$.
Using the effective K\"ahler potential Eq.~(\ref{eq:effkahlerp}) and the superpotential Eq.~(\ref{eq:effsuperp}), we obtain the potential of $S$ as,
\beq
V \sim (S^\dagger S)^{\frac{\D_S-1}{\D_S}-1+\frac{N_Q}{N_C}}.
\eeq
Thus, if $\D_S>N_C/N_Q$, the potential is an increasing function of $S$ in the large vev region and the runaway is stopped.

The above argument breaks down when the vev of $S$ becomes so small that the mass $m_P$ is not negligible.
In this region, the model becomes non-calculable, but the SUSY breaking has been established~\cite{Izawa:2009nz,Yanagida:2010wf}.
Furthermore, there are some indirect discussions~\cite{Yanagida:2010wf} that the vev of $F$-term of $S$, $\vev{F_S}$, is nonzero 
if the Seiberg dual of this model~\cite{Barnes:2004jj} is not weakly coupled. Also, there is no strong reason that the vev of the lowest
component of $S$ (which we denote by the same symbol as the chiral field itself) is zero. Therefore we assume $\vev{S} \neq 0$ 
in this paper~\footnote{More comments should be made on this point. If the model is weakly coupled in the dual magnetic theory, then
the model has an ISS-like SUSY breaking vacuum~\cite{Yanagida:2010wf,Amariti:2010sz}.
Then, one can show that $\vev{F_S}=0$ and also $\vev{S}=0$. 
One reason that we have $\vev{S}=0$ in the weakly coupled dual magnetic theory is that this point is an R-symmetry
enhancement point. It is often the case that a vacuum is at such a symmetry-enhancement point, since a derivative of the potential 
with respect to a charged field is zero at such a symmetric point.
On the other hand, if the electric theory is weakly coupled, one can expect $F_S \neq 0$~\cite{Yanagida:2010wf}. 
Then the continuous R-symmetry of the model is already broken by $F_S$ since $F_S$ has a non-vanishing R-charge. Then there remains a discrete R symmetry, and the problem is whether this discrete R symmetry is spontaneously broken or not.
}. However, as emphasized in Ref.~\cite{Yanagida:2010wf}, the assumption $\vev{S}\neq 0$ is not essential in the present model, and is made
only for simplicity of the following discussions.

Let us now discuss the gauge mediation. As mentioned above, we gauge the flavor symmetry $SU(N_P)$ of the quarks $P,{\tilde P}$ by $SU(5)_{\rm GUT}$.
Then, these quarks play a role of the messenger fields. Because of the strong dynamics, the calculation of mass spectrum of the MSSM particles is
very difficult, so we use an effective operator analysis~\cite{Ibe:2007wp}.
From a simple loop counting one can check that the lowest dimensional operator generating the sfermion masses may be estimated as
\beq
{\cal L}_{\rm eff}^{\rm sfermion} \sim C(\f_{\rm SM})\frac{1}{|m_{P}|^2}\left(\frac{g^2_{\rm SM}}{16\pi^2}\right)^2 \left( \frac{N_Cg_h}{16\pi^2}\right)^2 \int d^4\h \tr [(\l S)^\dagger (\l S)] \f_{\rm SM}^\dagger \f_{\rm SM} \label{eq:sfermion}
\eeq
where $\f_{\rm SM}$ is an MSSM chiral field, $g_{\rm SM}$ is the MSSM gauge coupling, $g_h$ is the hidden sector gauge coupling, and 
$C(\f_{\rm SM})$ is the quadratic Casimir invariant of  $\f_{\rm SM}$. We have neglected the Dynkin index of the SM gauge group.
The lowest dimensional operator generating the gaugino masses may be estimated as
\beq
{\cal L}_{\rm eff}^{\rm gaugino} \sim \frac{1}{|m_{P}|^6} \left(\frac{g^2_{\rm SM}}{16\pi^2}\right) \left( \frac{N_Cg^2_h}{16\pi^2} \right)^\ell \int d^4 \h \tr[ (\l S)^\dagger (\l S) (\l S)^\dagger D^2 (\l S)] W_{\rm SM}W_{\rm SM} \label{eq:gaugino}
\eeq
where $W_{SM}$ is an MSSM gauge field strength chiral field, and $\ell$ is a loop number to generate the operator which, according to Ref.~\cite{Ibe:2007wp},
is 4.~\footnote{Whatever the value of $\ell$ is, one may write down a planar diagram so as to avoid $1/N$ suppressions.}
The model is very strongly coupled~\cite{Yanagida:2010wf}, then using the rules of naive dimensional analysis~\cite{Luty:1997fk} with factors
of $N$'s included (and assuming $N_C,~N_Q $ and $N_P$ are all of the same order, say $N$),
we may have $|m_P| \sim \L$, $N_C g_h^2/16\pi^2 \sim 1$, and $\l S^{\tilde j}_{~i} \sim (\L+\L^2 \h^2) \d^{\tilde j}_{~i}$.
Then we obtain a very crude estimate for the sparticle masses,
\beq
m^2_{\rm sfermion} \sim C(\f_{\rm SM}) \left(\frac{g^2_{\rm SM}}{16\pi^2}\right)^2  N \L^2,~~~~~~m_{\rm gaugino} \sim \left(\frac{g^2_{\rm SM}}{16\pi^2}\right)  N \L.
\label{eq:GMsoftmasses}
\eeq

Finally, we comment on the color and flavor numbers $N_C,N_Q$ and $N_P$ (details can be found in Ref.~\cite{Yanagida:2010wf}). 
As discussed above, we take $N_P=5$ to identify
$P,\tilde{P}$ as the $\bf{5}+\bar{\bf 5}$ messenger fields. Then, we take $N_C \leq 4$ to avoid the Landau pole problem of 
the SM gauge coupling~\cite{Jones:2008ib,Sato:2009yt}
and $N_Q < N_C$ for the SUSY breaking mechanism to work. These requirements lead to $(N_C,N_Q,N_P)=(4,3,5)$ uniquely.
In this case, the coupling constants of the infrared fixed point are very large and hence the dynamical scale is the same order as the messenger mass.
We restrict the following discussion to the case $(N_C,N_Q,N_P)=(4,3,5)$.

\subsection{Dark matter}\label{sec:darkmatter}
In the model discussed above, we have some global symmetries. Most of them can be broken explicitly by adding interactions other than the ones
in Eq.~(\ref{eq:treesuper}). However, there is a $U(1)$ symmetry which is difficult to be broken explicitly~\cite{Hamaguchi:2007rb}. 
The $U(1)$ symmetry, which we call $U(1)_{h}$,
is defined by the charge assignment,
\beq
Q,~P:+1,~~~~~~~~\tilde{Q},~\tilde{P}:-1,
\eeq
and charge 0 to all other fields (including the MSSM fields). It is easy see that we have to contract $SU(N_C=4)$ gauge indices 
by using the totally anti-symmetric tensor of this gauge group to write down an gauge invariant operator which have nonzero $U(1)_h$ charge.
For example, we need at least four quarks among $Q$ and $P$ to write down a gauge invariant chiral field with positive $U(1)_h$ charge.
However, because there are only $N_Q=3$ flavors of quarks $Q$, and because $P$ quarks transform as ${\bf 5}$ under $SU(5)_{\rm GUT}$,
we cannot write down a gauge invariant chiral field which have mass dimension 4. Thus, to break $U(1)_h$
in a superpotential, we need dimension 6 or higher operators such as
\beq
\int d^2\h\frac{1}{M_{\rm UV}^2}QQQP \bar{\bf 5}_M,
\eeq
where 
$\bar{\bf 5}_M$ denotes the MSSM quarks and leptons in $\bar{\bf 5}$ representation of the $SU(5)_{\rm GUT}$ , and $M_{\rm UV}$ is some cutoff scale such as the GUT or the Planck scale.
To break $U(1)_h$ in a K\"ahler potential, we need dimension 6 or higher operators as, e.g.
\beq
\int d^4\h \frac{1}{M_{\rm UV}^2}QQ{\tilde Q}^\dagger {\tilde Q}^\dagger .
\eeq
We can identify the lightest $SU(4)$ gauge invariant composite state charged under $U(1)_h$ as the DM,
and this DM can be sufficiently long lived if $M_{\rm UV}$ is around the GUT or Planck scale~\cite{Hamaguchi:2007rb}.
However, the neutrality of the DM under the SM gauge interactions is not automatic in the present model.
If the DM candidate is composed only of $Q$ and $\tilde{Q}$, it is certainly neutral and a good candidate for the DM, but it is not easy to prove that
it is the case.
Because the calculation of the mass spectrum of the hidden sector particles is almost impossible, we simply assume the neutrality of the DM in this paper.

The annihilation cross section of the DM is close to the unitarity bound, since the DM is a bound state of strongly coupled confining force. 
As explained in section~\ref{sec:2} the dynamical scale $\L$ is required as $\L \sim {\cal O}(100)~\TEV$ to reproduce the observed value of the DM mass
density~\cite{Griest:1989wd}.

If we do not introduce other interactions, there are vector $U(1)$ symmetries conserved separately for $Q,\tilde{Q}$ and $P,\tilde{P}$.
Because $P, \tilde{P}$ are charged under the SM gauge group, we do not want the $U(1)_P$ symmetry which acts on $P,\tilde{P}$ to be 
conserved. To break $U(1)_P$ without affecting the low-energy physics, 
we may introduce couplings such as $W \sim Q\tilde{P} {\bf 10}_M \bar{\bf 5}_M /M_{\rm UV}$, where
${\bf 10}_M$ is the MSSM matter field in the ${\bf 10}$ representation of $SU(5)_{\rm GUT}$.
Then the lifetime of $P$ is estimated to be of order ${\cal O}(10^{-6})$~sec for $M_{\rm UV} \sim 10^{16}~\GEV$, 
and may decay before the period of the Big Bang Nucleosynthesis.

\section{Other issues of the model}

\subsection{$\mu$-$B_\mu$ }\label{sec:3.3}
There are many attempts to solve the so-called $\mu$ problem in gauge mediation.
In this subsection we present a possible solution to the problem, though it is not a strict requirement on the model.

We can write down the following couplings between the hidden sector fields and the up and down type Higgs fields:
\beq
W_{\rm Higgs}=(\l_s)^i_{~\tilde j} S^{\tilde j}_{~i} H_u H_d +(\l_u)_i  Q^i {\tilde P}_2 H_u+(\l_{d})^{\tilde j} {\tilde Q}_{\tilde j} P_2 H_d,\label{eq:higgshiddenint}
\eeq
where $P_2$ is the``${\bf 2}$ messengers'' transforming as ${\bf 2}_{\frac{1}{2}}$ under the decomposition ${\bf 5} \to {\bf 3}_{-\frac{1}{3}} +{\bf 2}_{\frac{1}{2}}$
of the representation of $SU(5)_{\rm GUT}$, and similarly for $\tilde{P}_2$. 
The first term generates $\mu$ and $B_\mu$ as
\beq
\mu \simeq \l_s \vev{S},~~~~~B_\mu \simeq \l_s \vev{F_S},\label{eq:treehiggs}
\eeq
with the vevs given as
\beq
 \vev{S} \sim \frac{\L}{4\pi},~~~~~\vev{F_S}\sim \frac{\L^2}{4\pi}. \label{eq:naiveSFS}
\eeq
The second and third terms may generate $\mu$ and $B_\mu$ at the one-loop level~\cite{Dvali:1996cu}
\beq
\mu \sim \frac{\l_u\l_d}{16\pi^2} \L,~~~~~B_\mu \sim \frac{\l_u\l_d}{16\pi^2}\L^2. \label{eq:loophiggs}
\eeq
Here 
$\L \sim m_P$ is the dynamical scale
of the model. See Ref.~\cite{Komargodski:2008ax} for a general discussion on the type of couplings Eq.~(\ref{eq:higgshiddenint}).
Here we have used the naive dimensional analysis (with the factors of $N$ neglected).

We assume that $\l_u,\l_d$ and $\l_s$ are not so large as to affect the infrared fixed point.
Then, the renormalization group equations at the fixed point show that these couplings run in the renormalization group as
\beq
\l_s |_{\L} \sim \left(\frac{\L}{M_*}\right)^{\g_S/2}\l_s |_{M_*},~~~~\l_{\{u,d\}} |_{\L} \sim \left(\frac{\L}{M_*}\right)^{(\g_Q+\g_P)/2}\l_{\{u,d\}} |_{M_*}, \label{eq:RGhiggs}
\eeq
where $M_*$ is the scale at which the theory enters into the infrared fixed point, and $\g_S,\g_Q$ and $\g_P$ are the anomalous dimensions 
of $S$, $Q,\tilde{Q}$ and $P,\tilde{P}$ at the fixed point, respectively. The subscripts $|_{\L}$ and $|_{M_*}$ mean that the couplings are evaluated
at the energy scale $\L$ and $M_*$, respectively. The values of the anomalous dimensions are given by
$\g_S \simeq 0.70,\g_Q \simeq -0.35,\g_P \simeq -0.59$ as shown in Ref.~\cite{Yanagida:2010wf}. Thus it is natural that $\l_s$ is much smaller than $\l_u$ and $\l_d$ at the energy scale $\L$,
and the tree-level contribution Eq.~(\ref{eq:treehiggs}) may not be too large compared with the one-loop suppressed contribution Eq.~(\ref{eq:loophiggs}).

All of the equations (\ref{eq:treehiggs}), (\ref{eq:naiveSFS}) and (\ref{eq:loophiggs}) suggest that the ratio of
$\mu$ and $B_\mu$ is of order
\beq
B_\mu /\mu \sim \L,
\eeq
with $\L \sim {\cal O}(100)~\TEV$. Thus, $\sqrt{B_\mu}$ is much larger than $\mu$.
This is the famous $\mu$-$B_\mu$ problem~\cite{Dvali:1996cu}. 
However, it has been pointed out in Ref.~\cite{Csaki:2008sr} that such a large $B_\mu$ is not necessarily inconsistent with the electroweak 
symmetry breaking conditions. The electroweak conditions are given by
\beq
\frac{m_Z^2}{2}=-|\mu|^2-\frac{m^2_{H_u}\tan^2\b-m^2_{H_d}}{\tan^2\b-1},~~~~~\frac{2\tan\b}{\tan^2\b+1}=\frac{2|B_\mu|}{2|\mu|^2+m^2_{H_u}+m^2_{H_d}}.\label{eq:EWcondition}
\eeq
These conditions can be satisfied if the mass parameters satisfy the following hierarchy,
\beq
|\mu|^2 \sim |m^2_{H_u}| \ll |B_\mu| \ll m^2_{H_d}. \label{eq:higgshierarchy}
\eeq
Then, Eq.~(\ref{eq:EWcondition}) can be solved (in the limit $\tan^2\b \gg 1$) roughly as $\tan\b \sim m^2_{H_d}/|B_\mu|$ and $m_Z^2/2 \sim -|\mu|^2-m^2_{H_u}+m^2_{H_d}/\tan^2\b$. 
The second and third terms in Eq.~(\ref{eq:higgshiddenint}) may give a contribution to the Higgs soft masses of order
\beq
m^2_{H_u} \sim \frac{|\l_u|^2}{16\pi^2}\L^2,~~~~~m^2_{H_d} \sim \frac{|\l_d|^2}{16\pi^2}\L^2,
\eeq
in addition to the gauge-mediation contribution Eq.~(\ref{eq:GMsoftmasses}).
Then, the hierarchy Eq.~(\ref{eq:higgshierarchy}) may be achieved for a large value of $\l_d$ and a small value of $\l_u$.
Note that in our model, such a large value of $\l_d$ is not unnatural
due to the renormalization group effect Eq.~(\ref{eq:RGhiggs}), while 
we have to require a somehow small value (or even zero) for $\l_u$ at the UV scale~\footnote{
To avoid a CP violation, it is perhaps necessary that one of the parameters $\l_u$ or $\l_s$ is negligibly small.
Then, we can rotate the phases of the Higgs fields to get real Yukawa couplings for $\l_d$ and either $\l_u$ or $\l_s$.
The phases of the hidden sector parameters can also be rotated away by phase rotations of the hidden sector fields,
so all the MSSM parameters can be taken real without loss of generality.
}.

\subsection{Discrete R symmetry}
There are many motivations to introduce discrete $R$ symmetry~\cite{Kurosawa:2001iq,Dine:2009swa}.
\begin{enumerate}
\item Discrete R symmetry is needed to ensure the SUSY breaking. In the case of our model,
there are gauge singlet fields $S^{\tilde j}_{~i}$, and the couplings of these fields cannot be restricted by gauge symmetries.
For the present SUSY breaking model to be successful, we need a specific form of the superpotential $W \sim SQ\tilde{Q}+P\tilde{P}$ in Eq.~(\ref{eq:treesuper}). 
Such a structure may be 
ensured by imposing a discrete R symmetry on the model.
\item In the previous subsection, we have discussed a dynamical generation of $\mu$ and $B_\mu$ terms. 
However, to really solve the $\mu$ problem,  we should forbid a tree level contribution to the $\mu$ term, which may in principle 
be of order $M_{\rm UV}$. 
\item The R parity in the MSSM is used to forbid dimension 4 proton decay interactions. However, there are dimension 5 operators
which may be quite dangerous for the proton decay problem, and cannot be forbidden by only the R parity.
\item To cancel the cosmological constant, the expectation value of the superpotential must be of order $\vev{W} \simeq m_{3/2}M_{Pl}^2$,
which is much smaller than $M_{UV}^3$. To forbid the contribution of order $M_{UV}^3$ by some symmetry, it should be an
R symmetry, since the superpotential is neutral under any non-R symmetries. However, if the R is a continuous symmetry, it predicts a
too small vev of the superpotential~\cite{Dine:2009sw}.
\end{enumerate}

\begin{table}[Hb]
\begin{center}
\begin{tabular}{|c|c|c|c|c|c|c|c|c|}
\hline
&$\bar{\bf 5}_M$&${\bf 10}_M$&$H_u$&$H_d$&$S$&$Q$&${\tilde Q}$&$P,{\tilde P}$                \\ \hline 
charge (mod 6)&$3$&$-1$&$-2$&$0$&$-2$&$3$&$1$&$1$\\ \hline
\end{tabular}
\caption{An example of discrete R symmetry. $\bar{\bf 5}_M$ and ${\bf 10}_M$ are the MSSM baryons and leptons written in the $SU(5)_{\rm GUT}$ representations.
The symmetry group is ${\mathbb Z}_{6}$, and the table shows the charge assignment
which is defined mod 6. The ${\mathbb Z}_2$ subgroup of this ${\mathbb Z}_6$ group can be seen as the usual R parity extended to the hidden sector.}
\label{table:1}
\end{center}
\end{table}

In this paper we do not discuss the issue of the cosmological constant problem, and only consider the first three motivations stated above.
In Table~\ref{table:1}, we show an example of discrete R symmetry ${\mathbb Z}_{6R}$ which can be imposed on our model.
We have required that the superpotentials Eq.~(\ref{eq:treesuper}), (\ref{eq:higgshiddenint}) and all the Yukawa couplings 
in the MSSM are allowed by the symmetry. As to the anomalies, we have required that the anomalies $[SU(2)_L]^2 {\mathbb Z}_{6R}$,
$[SU(3)_C]^2 {\mathbb Z}_{6R}$ and $[SU(4)]^2{\mathbb Z}_{6R}$ are cancelled. 
Anomaly cancellation is not a necessary condition, because there may be some fields which contribute to the anomaly and 
get masses from the spontaneous breaking of the discrete R symmetry (see~\cite{Kurosawa:2001iq,Dine:2009swa} and references therein).
However, in low-scale gauge mediation with many messenger fields, 
we do not want to have too many fields charged under the SM gauge group because of the Landau pole problem.
Thus, it is desirable that the SM gauge group anomalies are cancelled within the MSSM+messenger sector.

In the hidden sector, ${\mathbb Z}_{6R}$ forbids all the unwanted renormalizable interactions other than a mass term of $S$,
$m_SS^2$. We simply assume that $m_S$ is much smaller than $m_P$ so that this term does not affect the SUSY breaking.
Note that if $m_S$ and $m_P$ are the same order at the UV scale, then $m_S \ll m_P$ is naturally achieved by the renormalization group effect 
in the hidden sector. In the MSSM sector, a tree level $\mu$ term is forbidden by ${\mathbb Z}_{6R}$. The dimension 5 operators respecting
the usual R parity (which is the ${\mathbb Z}_2$ subgroup of ${\mathbb Z}_{6R}$) are given by
\beq
W &\sim& ({\bf 5}'_H \bar{\bf 5}_M)^2+{\bf 10}_M {\bf 10}_M {\bf 10}_M \bar{\bf 5}_M, \label{eq:dim5op}\\
K &\sim& {\bf 5}'_H {\bf 10}_M^\dagger \bar{\bf 5}_M^\dagger+\bar{\bf 5}'_H {\bf 10}_M^\dagger {\bf 10}_M^\dagger +{\rm h.c.},
\eeq
where we have defined ${\bf 5}'_H =(0,H_u)$ and $\bar{\bf 5}'_H=(0,H_d)$. Among them, only the first term in Eq.~(\ref{eq:dim5op}) 
(which gives masses to neutrinos) is allowed by the ${\mathbb Z}_{6R}$ symmetry, and the second term in Eq.~(\ref{eq:dim5op}) 
(which is a dangerous dimension 5 operator for the proton decay) is forbidden. 
One can also check that there are no dimension 5 baryon number violating operators even including the hidden sector. 
The coupling $W \sim \bar{\bf 5}_M P\tilde{Q}$ is also not allowed which may be dangerous for the flavor violation.
Thus, the ${\mathbb Z}_{6R}$ symmetry of Table~\ref{table:1} is very attractive, but it is shown only for the purpose of illustration,
and there may be other symmetries or mechanisms to solve the problems stated above~\footnote{
Unfortunately, the discrete R-symmetry forbids the operator $W \sim Q\tilde{P} {\bf 10}_M \bar{\bf 5}_M $ discussed in subsection~\ref{sec:darkmatter}.
Although $P_2,\tilde{P}_2$ decays to $Q,\tilde{Q}$
 through Eqs.~(\ref{eq:higgshiddenint}), the decay of the quarks $P_3,\tilde{P}_3$ becomes somehow difficult.
The decay can be induced by e.g. introducing new quarks $T, \tilde{T}$ in the fundamental and anti-fundamental representations of $SU(4)$,
with a superpotential $W \sim \bar{\bf 5}_M P\tilde{T}+m_T T \tilde{T}~~(m_T \gg m_P)$.   
Then, $P_3,\tilde{P}_3$ decay to (on-shell or off-shell) $P_2,\tilde{P}_2$ and other MSSM particles.
}.

The ${\mathbb Z}_{6R}$ symmetry is broken spontaneously by the vev of $S$, so we should care about domain walls.
If ${\mathbb Z}_{6R}$ acts also on the constant term, 
it is broken at some high-energy scale to generate the constant term, and the domain wall may be inflated away. See Ref.~\cite{Dine:2010eb} for details.

\subsection{Mass spectrum and LHC signature}
It is difficult to calculate the MSSM mass spectrum due to the strong coupling, but we may infer some features of the model.
First, colored particles are most likely lighter in the present model compared with in the minimal gauge mediation~\cite{Ibe:2007wp}.
The mass of the $P_3,\tilde{P}_3$ messengers, $m_{P3}$, and the mass of the $P_2,\tilde{P}_2$ messengers, $m_{P2}$, 
are different in general. 
We assume that $m_{P3}=m_{P2}$ at the GUT scale, but these masses run differently under the renormalization group.
$SU(3)_C$ interactions make $m_{P3}$ slightly larger than $m_{P2}$. Then messenger loops involving $P_3,\tilde{P}_3$ are slightly suppressed.
If the operators Eq.~(\ref{eq:sfermion}) and (\ref{eq:gaugino}) give the dominant contribution to the soft masses, then
the colored particle mass dependence on $r \equiv m_{P3}/m_{P2}$ is estimated as $M_{\rm gluino} \propto r^{-6}$ and $m_{\rm squark} \propto r^{-1}$.
See Ref.~\cite{Hamaguchi:2008yu,Shirai:2010rr} for details. 
However, in the present model the SUSY breaking scale is comparable to the messenger mass scale and higher dimensional operators 
could contribute to the soft masses, and hence the precise dependence on $r$ is not obvious.

Second, the gauginos may be lighter than the sfermions. At least in the case of weakly coupled calculable semi-direct gauge mediation models,
there are numerical suppressions
in the gaugino masses~\cite{Shirai:2010rr}. Although the present model is not calculable at all , it is presumable that the gauginos are somehow lighter.

The above arguments suggest that
the next-to-lightest SUSY particle (NLSP) is 
perhaps the Bino or the gluino~\cite{Hamaguchi:2008yu,Shirai:2010rr}.
If the Bino is the NLSP, then the situation is very similar to the case studied in Ref.~\cite{Sato:2010tz} (see also Refs.~\cite{Feng:2010ij},\cite{Shirai:2009kn}),
and it may be possible to discover SUSY events at early stages of the LHC with $7~\TEV$ run.
On the other hand, if the gluino is the NLSP, then the gluino pair is produced at the LHC and then decays to gluons and gravitinos, 
thus the typical SUSY signal is di-jet+missing energy. 

An R-axion may be observed at the LHC~\cite{Goh:2008xz}. The hidden sector model has an approxiate continuous R symmetry with charge assignment
$S:\frac{8}{3},~Q,\tilde{Q}:-\frac{1}{3},~P,\tilde{P}:1$. This R symmetry is respected by the MSSM sector up to MSSM anomalies
if either $\l_u$ or $\l_s$ is negligibly small as mentioned in subsection~\ref{sec:3.3}.
If one uses the naive dimensional analysis, the decay constant of the R-axion is of order $f_a \sim \L/4\pi \sim {\cal O}(10)~\TEV$.
This is in the region of the parameter space studied in Ref.~\cite{Goh:2008xz}, and the R-axion detection may be possible by
searching displaced vertices in the decay of an R-axion to two muons.

\section*{Acknowledgements}
We would like to thank K.~Hamaguchi, E.~Nakamura, R.~Sato and S.~Shirai for useful discussions.
The work of KY is supported in part by JSPS Research Fellowships for Young Scientists.
This work was supported by World Premier International Research Center Initiative (WPI Initiative), MEXT, Japan.


\end{document}